\documentclass[twocolumn,trackchanges]{aastex701}
\RequirePackage{lineno}
\usepackage{amsmath}
\usepackage{amssymb}
\usepackage{amsthm}
\usepackage{natbib,aasdefs,url,bm}
\usepackage{array}
\usepackage{float}
\usepackage{lipsum,graphicx}
\usepackage{subfigure}
\usepackage{color}
\newcounter{ichi}
\newcounter{ni}
\newcounter{san}
\newcounter{yon}
\def\be{\begin{equation}}
\def\ee{\end{equation}}
\def\ba{\begin{eqnarray}}
\def\ea{\end{eqnarray}}

\newcommand{\hide}[1]{}
\def\aap{\ {A\&A}\ }

\def\apjl{\ {ApJL}\ }
\def\apjs{\ {ApJS}\ }

\shorttitle{Multimessenger Constraints on Production Sites of NGC 1068 Neutrinos}
\shortauthors{Das, Murase and Zhang}

\begin{document}

\title{Multimessenger Constraints on Production Sites of High-Energy Neutrinos from NGC 1068}%\vspace{-5em}}

\author[orcid=0000-0003-2916-0211]{Abhishek Das}
\affiliation{Department of Physics, The Pennsylvania State University, University Park, PA 16802, USA}
%\affiliation{Department of Astronomy \& Astrophysics, The Pennsylvania State University, University Park, PA 16802, USA}
\affiliation{Institute for Gravitation and the Cosmos, The Pennsylvania State University, University Park, PA 16802, USA}
\email{ajd6518@psu.edu}

\author[orcid=0000-0002-5358-5642]{Kohta Murase}
\affiliation{Department of Physics, The Pennsylvania State University, University Park, PA 16802, USA}
\affiliation{Department of Astronomy \& Astrophysics, The Pennsylvania State University, University Park, PA 16802, USA}
\affiliation{Institute for Gravitation and the Cosmos, The Pennsylvania State University, University Park, PA 16802, USA}
\affiliation{Center for Gravitational Physics and Quantum Information, Yukawa Institute for Theoretical Physics, Kyoto University, Kyoto 606-8502, Japan}
\email{murase@psu.edu}

\author[orcid=0000-0003-2478-333X]{B. Theodore Zhang} 
\affiliation{Key Laboratory of Particle Astrophysics and Experimental Physics Division and Computing Center, Institute of High Energy Physics, Chinese Academy of Sciences, 100049 Beijing, China}
\affiliation{TIANFU Cosmic Ray Research Center, 610213,Chengdu, Sichuan , China}
\email{zhangbing@ihep.ac.cn}

\begin{abstract}
The detection of high-energy neutrino signals from the nearby Seyfert galaxy NGC 1068 provides us with a unique opportunity to explore nonthermal processes near the center of supermassive black holes. Using the IceCube and {\it Fermi}-LAT data, we present general multimessenger constraints on the energetics of cosmic rays and the extent of the neutrino emission region, considering not only photohadronic (photomeson and photopair production) but also hadronuclear ($pp$) processes. Compared to the photohadronic scenario, the hadronuclear scenario can alleviate constraints on the emission region, yielding $R\lesssim30-70R_S$ for low-$\beta$ plasma, where the magnetic pressure dominates the gas pressure, and $R\lesssim5-50R_S$ for high-$\beta$ plasma. While our results support the previous conclusion that the photohadronic scenario favors a compact corona with $R\sim3-10R_S$, these suggest the relevance of further investigations into $pp$ neutrino contributions. When the the cosmic-ray spectrum is extended down to a minimum energy of 1~GeV, which is natural in light of particle acceleration mechanisms, we find that the required cosmic-ray luminosity exceeds the X-ray luminosity for a spectral index of $s_{\rm CR}\gtrsim2$, which challenges some shock acceleration models. We also show that the beta decay scenario is unlikely even if the magnetic field is as strong as the maximum allowed by the Eddington luminosity. Given that NGC 1068 can be established as a neutrino source, our results will provide evidence for the standard hadronic scenario, including magnetically powered corona models having hard spectra with $s_{\rm CR}\lesssim2$. 
\end{abstract}


\section{Introduction}

\renewcommand{\thefootnote}{\arabic{footnote}}
\setcounter{footnote}{0}

The origin of high-energy cosmic rays and neutrinos has remained an enigma since their discovery by the IceCube Collaboration ~\citep{Aartsen:2013bka, Aartsen:2013jdh}. While inelastic hadronuclear ($pp$) and/or photomeson production ($p\gamma$) processes that generate neutrinos should also produce gamma rays of similar energies, multimessenger measurements of the all-sky neutrino and gamma-ray fluxes indicate that the sources responsible for the all-sky neutrino background are hidden or opaque for GeV--TeV gamma rays
~\citep{Murase:2013rfa,Murase:2015xka,Capanema:2020rjj,Fang:2022trf}. This is because the non-blazar component of the isotropic diffuse gamma-ray background is violated if all neutrino sources are gamma-ray transparent. In 2022, the observation of $\sim1-10$~TeV neutrinos from the nearby Seyfert galaxy, NGC 1068~\citep{IceCube:2022der} with a significance of $\sim4\sigma$, provided a unique opportunity to address the mystery of neutrino production. 
The neutrino luminosity inferred from the IceCube data for NGC 1068 is about 2 orders of magnitude larger than the observed gamma-ray luminosity of NGC 1068 in the {\it Fermi}-LAT GeV band as well as the upper limits in the TeV band~\citep{MAGIC:2019fvw,Fermi-LAT:2019yla,Fermi-LAT:2019pir,Ajello:2023hkh}, which makes this source a hidden neutrino-active galaxy~\citep{Murase:2022dog}. In recent years, neutrino excesses have also been identified from other active galactic nuclei (AGNs) such as NGC 4151~\citep{IceCube:2024ayt, IceCube:2024dou} and the Circinus Galaxy~\citep{IceCube:2026hzq}, and further searches have been performed~\citep{IceCube:2024fxo,IceCube:2025gdd}. 

The mechanism for particle acceleration and high-energy neutrino production in the vicinity of supermassive black holes (SMBHs) is currently uncertain and of interest. ``Jet-quiet" AGNs like NGC 1068 that do not have powerful jets are believed to have a spectral energy density (SED) featuring a ``big blue bump" that is an optical/UV component originating from multitemperature blackbody emission from an accretion disk, as well as power-law X-ray emission from a hot plasma region called a corona. Particle acceleration in the standard corona may naturally occur in magnetically powered, turbulent coronae\footnote{\cite{Murase:2019vdl} consider a phenomenological turbulent corona model applicable to both large-amplitude and small-amplitude turbulence~\cite[see][]{Murase:2011cx}. The framework has not been specified for weak turbulence~\citep{Testagrossa:2026jcs}, as also demonstrated in~\cite{Murase:2026hrz}.}, where particles are accelerated by turbulence and/or magnetic reconnections in low-$\beta$ plasma, where $\beta$ is the pressure ratio of gas to magnetic fields~\citep{Murase:2019vdl,Kheirandish:2021wkm,Eichmann:2022lxh,Fiorillo:2023dts,Lemoine:2024roa}. Alternatively, protons might be accelerated by shocks that could be caused by the free fall of the bulk material~\citep{Inoue:2019fil}, failed winds~\citep{Inoue:2022yak}, or reconnection-driven flows~\citep{Murase:2022dog}. Furthermore, particles could also experience shear acceleration by weak jets or outflows~\citep{Murase:2022dog,Lemoine:2024roa}.
Recent improved measurements of the all-flavor astrophysical neutrino spectrum from IceCube~\citep{IceCube:2024fxo,IceCube:2025tgp} have provided evidence for a break in the neutrino spectrum, which is consistent with a prediction of the magnetically powered AGN corona model~\citep{Murase:2026hrz}. 

\cite{Das_2024} explored multimessenger constraints on the photohadronic, leptonic, and beta decay scenarios. In the $p\gamma$ scenario, the size of the emission region is found to be $R\lesssim 15 R_S$, where $R_S$ is the Schwarzschild radius, which is tighter than the previously derived limit, $R\lesssim 30~R_S$~\citep{Murase:2022dog}. It has also been shown that the leptonic and beta decay scenarios are excluded as the dominant mechanism for the observed neutrinos from NGC 1068.

In this work, we extend the previous analysis by exploring the hadronuclear ($pp$) scenario for neutrino production, and present its implications for cosmic-ray energetics and multimessenger constraints on the neutrino emission region. The hadronuclear scenario has been previously explored in the literature~\citep{Murase:2019vdl,Murase:2022dog, Blanco:2023dfp}, and this work comprehensively investigates parameter space of the emission radius ($R$), the cosmic-ray spectral index ($s_{\rm CR}$), and the magnetic field.  In particular, \cite{Murase:2019vdl} pointed out the critical importance of the Bethe-Heitler process (as demonstrated in Appendix \ref{app:bh_importance}) although it is often ignored in the literature \citep[e.g.,][]{Blanco:2023dfp}. We evaluate the cosmic-ray luminosity considering this process self-consistently. We also revisit the beta decay scenario and present results for the strongly magnetized regime more explicitly. We use $\mathcal{R}\equiv R/R_S$ as the dimensionless emission radius expressed in units of Schwarzschild radius. Any quantity expressed as $\mathcal{Q}_x$ is defined as $\mathcal{Q}_x\equiv\mathcal{Q}/10^x$ in CGS units.

\section{Setup and Method of Calculations}

\begin{figure}
    \centering
    \includegraphics[width=\linewidth]{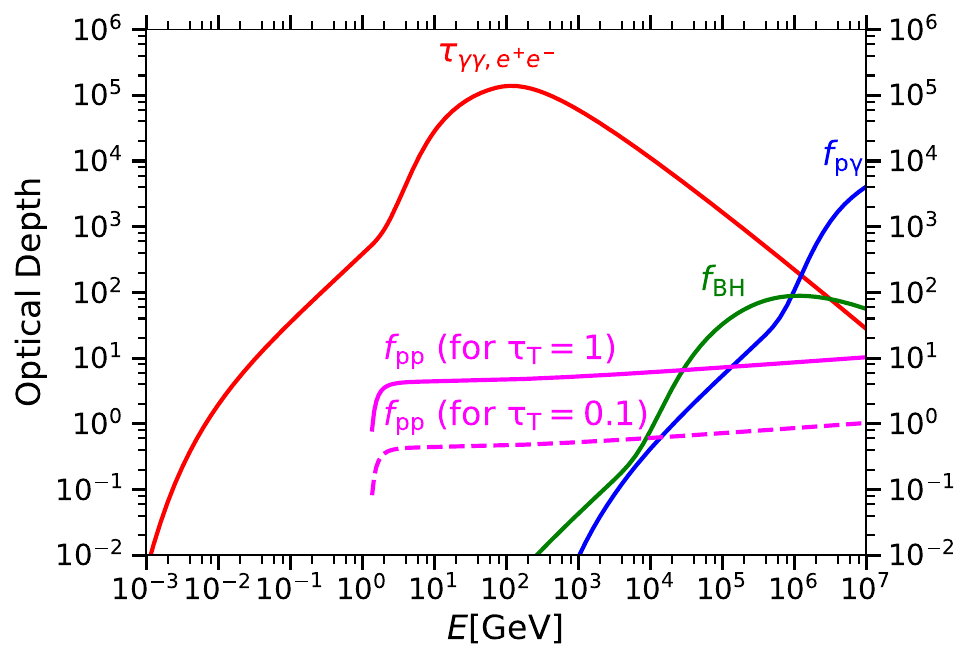}
    \caption{Optical depth for the two-photon annihilation ($\tau_{\gamma\gamma \rm,e^{+}e^{-}}$) and effective optical depths for photomeson production ($f_{p\gamma}$), Bethe-Heitler pair production ($f_{\rm BH}$), and inelastic proton-proton collisions ($f_{pp}$) for two values of $\tau_T$. The emission radius is set to $R = 10 R_S$. 
    For the photomeson production, the Bethe-Heitler process, and proton-proton collisions, the proton escape timescale is set to $t_* = 100 R/c$.
    }
    \label{fig:optical-depth}
\end{figure}

NGC 1068 is an archetypal Seyfert II galaxy at a luminosity distance of $d_L = 10\rm~Mpc$~\citep{Tully:2009ir,Courtois:2013yfa,2021AstBu..76..255T}.
We use $M_{\rm BH}=10^7~\text{M}_\odot$ as the SMBH mass~\citep{LodatoBertin,2020MNRAS.497.1020W}, although the self-gravity of the disk may lead to a lower mass of $M_{\rm BH}=6\times10^6~\text{M}_\odot$~\citep[see also][]{2002ApJ...579..530W,Panessa:2006sg,LodatoBertin}. The Eddington luminosity is $L_{\rm Edd}\simeq 1.3\times{10}^{45}~{\rm erg}~{\rm s}^{-1}~\left(M_{\rm BH}/10^7~M_\odot\right)$. NGC 1068 is a near-Eddington source, where the bolometric luminosity ($L_{\rm bol}$) is comparable to $L_{\rm Edd}$ and the Eddington ratio ($\lambda_{\rm Edd} = L_{\rm bol}/L_{\rm Edd}$) is $\sim 1$. 
For NGC 1068, the intrinsic $2-10$~keV X-ray luminosity is measured to be $L_X^{2-10~\rm keV}\approx3.4\times10^{43}~{\rm erg~s}^{-1}$~\citep{Marinucci:2015fqo}, and $L_{\rm X}$ is typically expected to be $\sim 20-30$ times lower than $L_{\rm bol}$.
Because NGC 1068 is a jet-quiet AGN, the cosmic-ray luminosity ($L_{\rm CR}$) is expected to be lower than $L_{\rm bol}$, and it is more reasonable to expect that the nonthermal luminosity is lower than the total X-ray luminosity, $L_{\rm corona}\approx7.8\times10^{43}~{\rm erg~s}^{-1}$~\citep{Murase:2015xka}. The cosmic-ray loading factor against the intrinsic X-ray luminosity, $\xi_{\rm CR/X}$, is introduced as $L_{\rm CR}\equiv\xi_{\rm CR/X}L_X$~\citep{Murase:2019vdl}. 
The magnetically powered corona model for explaining the all-sky neutrino flux 
suggests $\xi_{\rm CR/X}<1$ with a spectral index of $s_{\rm CR}\lesssim2$~\citep{Murase:2019vdl,Kheirandish:2021wkm,Murase:2026hrz}. The accretion shock model accounting for X rays by shock-heated plasma also expects $\xi_{\rm CR/X}<1$ but with $s_{\rm CR}\gtrsim2$~\citep{Inoue:2019fil}. 
The corona is energized through magnetic dissipation in the magnetically powered corona model, whereas the bulk kinetic energy of the free-fall material dissipates via shocks in the original accretion shock model~\citep{Meszaros:1983ed}.    
%Last but not the least, we should consider the acceleration fraction ($\epsilon_{\rm CR}$) which relates the required minimum cosmic ray luminosity to the dissipated power ($L_{\rm diss}$) as $L_{\rm CR}=\epsilon_{\rm CR}L_{\rm diss}$. Note that $L_{\rm diss}$ is different for different models. For corona models, $L_{\rm dis}$ is the magnetic dissipation power $L_B$, while in shock models, it is the kinetic dissipation power $L_{\rm kin}$. For the corona model with stochastic acceleration, $\epsilon_{\rm CR}\sim9\times10^{-4}-3\times10^{-2}$ for moderate to high cosmic ray pressure, and for reconnection $\epsilon_{\rm CR}\sim10^{-2}$~\citep{Kheirandish:2021wkm}. For shock-accelerated models, we have much larger values for $\epsilon_{\rm CR}\sim0.1-0.2$~\citep{Inoue:2022yak}. Thus, we can see that the corona models are far more robust than shock-accelerated models.

As in~\cite{Das_2024}, we use the {Astrophysical Multimessenger Emission Simulator (AMES)} to simulate the photohadronic, photonuclear and hadronuclear interactions in order to generate cascaded gamma-ray and neutrino spectra, and find out the required cosmic-ray luminosity $L_{\rm CR}$ for different scenarios, as well as the allowed regions of parameter space based on energetics and constraints from {\it Fermi}-LAT~\citep{Ajello:2023hkh}, MAGIC~\citep{MAGIC:2019fvw}, and IceCube~\citep{IceCube:2022der} data. 
We focus on the single emission zone model characterized by the coronal compactness ${\mathcal R}$. For details on the target photon spectrum, we refer to Section 2 of~\cite{Das_2024}. The magnetic field is parameterized in terms of $\xi_B \equiv U_B/U_\gamma$, where $U_B = B^2/8\pi$ is the magnetic energy density and $U_\gamma = L_\text{bol}/(4\pi R^2 c)$ is the bolometric photon energy density. Given that $L_{\rm bol}$ and $M_{\rm BH}$ are estimated from the observation, the magnetic field is converted from $\xi_B$ into 
\begin{align}
    B&=\left(\frac{2\xi_BL_{\rm bol}}{R^2 c}\right)^{1/2} \nonumber \\
    &\simeq9.7\times10^3~{\rm G}~\xi_B^{1/2}L_{\rm bol,45.1}M_{\rm BH,7}^{-1}\mathcal{R}_1^{-1}.
    \label{eq:xiB}
\end{align}
For given values of $\xi_B$ (that are characterized by models), the magnetic field strength is inversely proportional to the emission radius. This is the commonly used parameterization that is similar to so-called equipartition parameters. For example, the accretion and wind shock models typically predict $\xi_B\ll 0.1-1$~\citep{Murase:2022dog}. We note that in our setup pion and muon cooling effects do not have significant impacts on our results~\citep[see also][]{Blanco:2025zqo}. 

We also take into account gamma-ray attenuation in matter, as in~\cite{Murase:2022dog}, for which we use a column density of $N_{\rm H}=10^{25}~{\rm cm}^{-2}$~\citep{Marinucci:2015fqo}. 

Above TeV energies, gamma rays are further attenuated by infrared emission from the dust torus, whose characteristic radius is set to $R_\text{DT}=0.1$~pc~\citep{torus_ALMA1,torus_ALMA2}. The dust torus spectrum is approximated by a blackbody with a temperature of $T_\text{DT} = 1000$~K~\citep{Rosas:2021zbx,Inoue:2022yak} as $\varepsilon_\gamma(dn_\text{DT}/d\varepsilon_\gamma)=(8\pi/c^3h^3)\varepsilon_\gamma^3/(\exp(\varepsilon_\gamma/[k_B T_\text{DT}])-1)$, and 
we multiply $\exp(-\tau_{\gamma\gamma}^{\rm DT})$, where $\tau_{\gamma\gamma}^{\rm DT} = t_{\gamma \gamma}^{-1} R_{\rm DT}/c$ and $t_{\gamma\gamma}$ is the $\gamma \gamma$ interaction timescale in the infrared photon field. 
In the beta decay scenario, we also consider $R^{-2}$ dependence for $R>R_{\rm DT}$, where the target photon field is a combination of infrared photons from the dust torus, the cosmic microwave background (CMB), and extragalactic background light (EBL). We model the EBL using the model of~\cite{Gilmore:2011ks}, and account for gamma-ray attenuation due to CMB and EBL during intergalactic propagation. Note that the energies of the photons making up this component are well below the threshold for the production of neutrinos of interest; therefore, our results on neutrinos are unaffected by the implementation of this component. 

We assume a power-law distribution with an exponential cutoff for the injected cosmic-ray proton energy spectrum,
\begin{equation}
    \varepsilon_p L_{\varepsilon_p}\equiv \varepsilon_p \frac{dL_{\rm CR}}{d\varepsilon_p} \propto \varepsilon_p^{2-s_{\rm CR}} {\rm exp}\left({-\frac{\varepsilon_p}{\varepsilon_p^\text{max}}}\right),
    \label{eq:diffProtonDens}
\end{equation}
where $s_{\rm CR}$ is the proton power-law index and $\varepsilon_p^\text{max}$ is the maximum proton energy. The normalization is set by the cosmic-ray proton luminosity which is
\label{eq:injCR}
\begin{equation}
    L_{\rm CR}=\mathcal{C}_p(s_{\rm CR},\varepsilon_{p}^{0}) \times 
    (\varepsilon_p L_{\varepsilon_p})\rvert_{\varepsilon_{p}^{0}}
\end{equation}
where $\varepsilon_{p}^{0}$ is the reference energy for normalization, and $\mathcal{C}_p$ is the conversion factor from differential to bolometric luminosity~\citep{Murase:2022dog} which can be derived in terms of the incomplete Gamma function as follows.
\begin{eqnarray}
\mathcal{C}_p(s_{\rm CR})&=&\exp\left(\frac{\varepsilon_p^0}{\varepsilon_p^{\rm max}}\right)\left(\frac{\varepsilon_p^{\rm max}}{\varepsilon_p^0}\right)^{2-s_{\rm CR}} \nonumber \\
   &~&\times\Gamma\left(2-s_{\rm CR},\frac{\varepsilon_p^{\rm min}}{\varepsilon_p^{\rm max}}\right)
    \label{eq:Cp}
\end{eqnarray}

\begin{deluxetable}{ccccc}
\tablecaption{Conversion factor $\mathcal{C}_p$ from differential to bolometric luminosity for different spectral indices $s_{\rm CR}$, assuming $\varepsilon_p^0 =10~\mathrm{TeV}$. Two cases of $\varepsilon_p^{\rm min}$ are shown for spectra with an exponential cutoff or an abrupt cutoff at $\varepsilon_p^{\rm max}$. \label{tab:Cp}}
\tablehead{
\colhead{$s_{\rm CR}$} & 
\multicolumn{2}{c}{$\varepsilon_p^{\rm min} = 1~\mathrm{GeV}$} &
\multicolumn{2}{c}{$\varepsilon_p^{\rm min} = 10~\mathrm{TeV}$} \\
\colhead{} & 
\colhead{abrupt} & \colhead{exponential} &
\colhead{abrupt} & \colhead{exponential} 
}
\startdata
1.0 & $3.0$ & $4.2$ & $2.0$ & $3.0$ \\
1.5 & $3.4$ & $4.3$ & $1.5$ & $1.8$ \\
2.0 & $1.0 \times 10^{1}$ & $1.4 \times 10^{1}$ & $1.1$ & $1.2$ \\
2.5 & $2.0 \times 10^{2}$ & $2.8 \times 10^{2}$ & $0.85$ & $0.82$ \\
3.2 & $5.3 \times 10^{4}$ & $7.3 \times 10^{4}$ & $0.61$ & $0.56$ \\
\enddata
\end{deluxetable}

Table~\ref{tab:Cp} shows the values of $\mathcal{C}_p$ for different values of $s_{\rm CR}$ for both proton energy bands used in the hadronic scenario. Note that the abrupt cutoff case assumes a power-law spectrum truncated at $\varepsilon_p^{\rm max}$ while the exponential cutoff case assumes the form of the injected spectrum shown in Equation~\ref{eq:diffProtonDens} and used in our calculations.

In this work, we explore the energetics and constraints for hadronic scenarios for a narrow injected cosmic-ray spectrum ($\varepsilon_p^\text{min}=10$~TeV and $\varepsilon_p^\text{max}=30$~TeV) which is the same as what is used for the photohadronic scenario in~\cite{Das_2024}, which should effectively be regarded as the \textit{minimal hadronic scenario}. This scenario should be regarded as a conservative benchmark that minimizes the required bolometric cosmic-ray luminosity and associated cascade emission. It approximates hard or low-energy-suppressed spectra and is not meant to imply a literal sharp cutoff in the physical proton distribution. In addition, we explore the results for a broad injected cosmic-ray spectrum with $\varepsilon_p^\text{min}=1$~GeV while keeping $\varepsilon_p^\text{max}=30$~TeV, both for the photohadronic scenario and the hadronuclear scenario with the Thomson optical depth $\tau_T = 1$. When~GeV cosmic rays are considered, we set the dynamical velocity to $V=0.1 \rm c$, which is the same as what is used for the photohadronic scenario in~\cite{Das_2024}. We note that this velocity is also comparable to the free-fall velocity, $V_{\rm ff}\approx V_K$, where $V_K$ is the Keplerian velocity~\citep{Bondi52,Meszaros:1983ed}, and it is motivated by some of the shock models~\citep{Inoue:2019fil,Inoue:2022yak}. For the minimal hadronic scenario, we use $V=0.01 \rm c$, which is close to the infall velocity ($V_{\rm fall}\approx\alpha V_K$)~\citep[e.g.,][]{Shakura:1972te,NarayanYi94,NarayanYi95} where $\alpha$ is the viscosity parameter, and this is motivated by the disk-corona model~\citep{Murase:2019vdl,Kheirandish:2021wkm}.

\section{Standard Hadronic Scenario}
\label{sec:pp}

\begin{figure*}
    \centering
    \includegraphics[width=\linewidth]{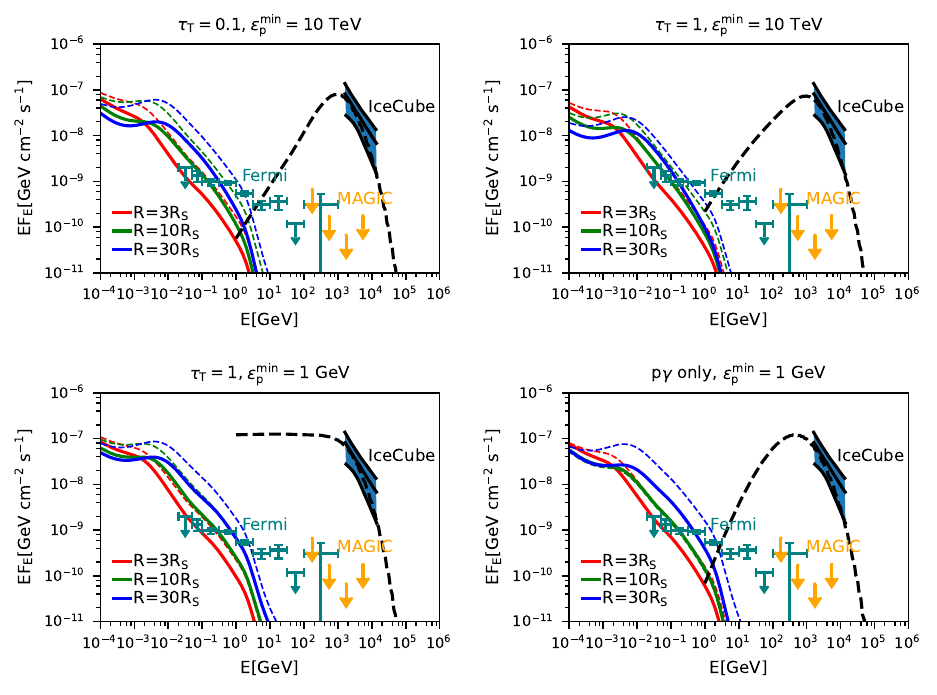}
    \caption{Cascaded photon spectra (colored lines) and all-flavor neutrino spectra (dashed black lines) for different values of the emission radius $R$ for different parameter sets explored in this work. Among the photon spectra, solid lines correspond to $\xi_B=1$ while dashed lines correspond to $\xi_B=0.01$. All panels are made for injected cosmic-ray protons with $s_{\rm CR}=2$ and $\varepsilon_{p}^{\rm max}=30$~TeV. The solid black lines correspond to the 95 percent contour lines and the best-fit line from the IceCube data~\citep{IceCube:2022der}. Gamma-ray data from the {\it Fermi}-LAT~\citep{Ajello:2023hkh} and MAGIC~\citep{MAGIC:2019fvw} observations are also shown.}
    \label{fig:spectrum-plots}
\end{figure*}

\begin{figure*}
    \centering
    \includegraphics[width=\linewidth]{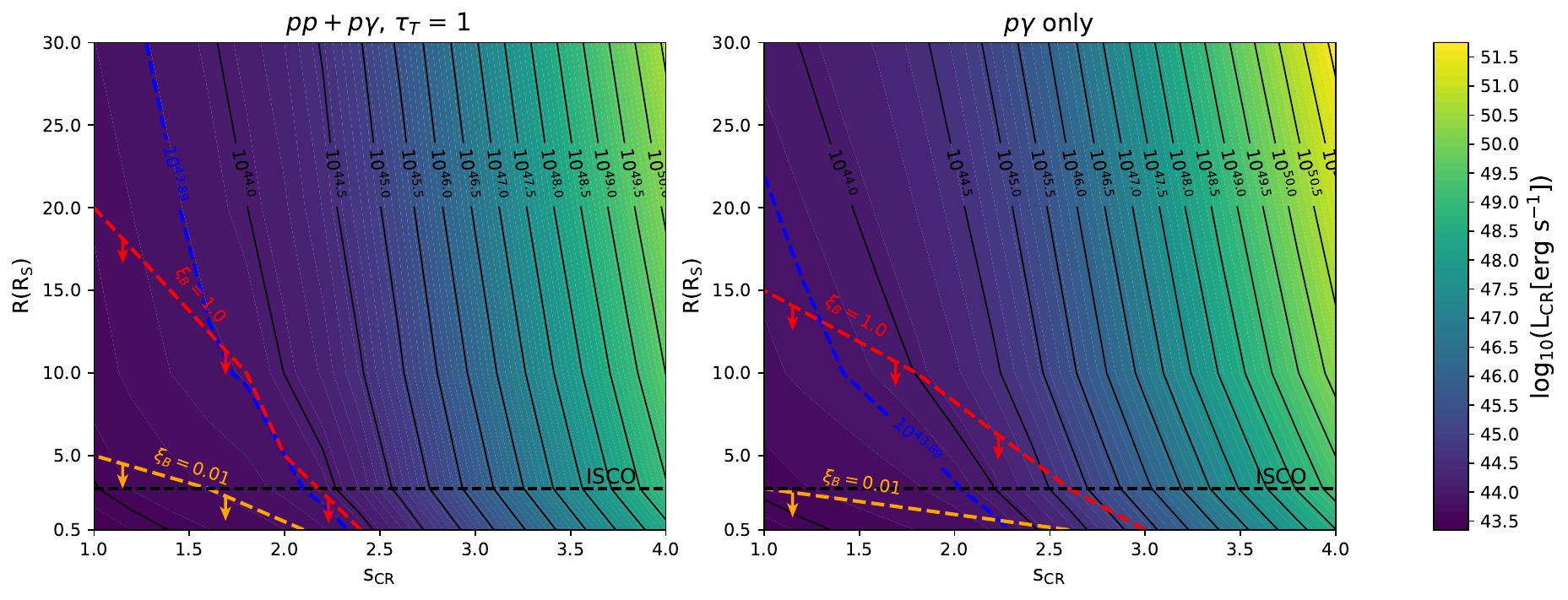}
    \caption{Required minimum cosmic-ray proton luminosity, $L_\text{CR}$, as a function of the emission radius $R$ and the power-law index $s_{\rm CR}$ for $\varepsilon_p^{\rm min}=1$~GeV and $\varepsilon_p^{\rm max}=30$~TeV in the hadronic scenario. The blue contour corresponds to $L_{\rm corona}$, which is the total coronal X-ray luminosity. The red and orange lines correspond to the upper limits from gamma-ray observations for $\xi_B=1$ and $\xi_B=0.01$, respectively.}
    \label{fig:contour-1gev}
\end{figure*}

\begin{figure*}
    \centering
    \includegraphics[width=\linewidth]{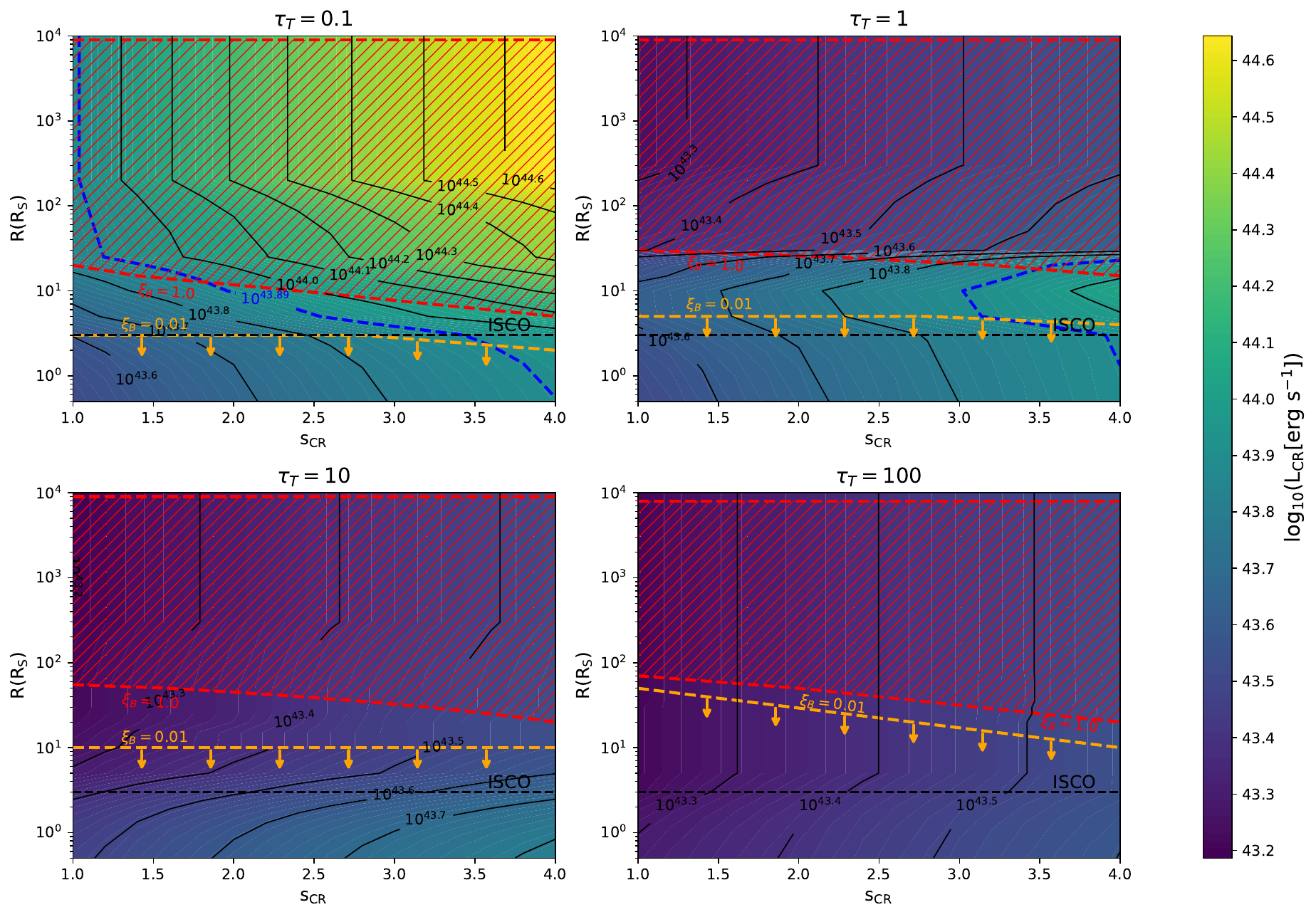}
    \caption{Required minimum cosmic-ray proton luminosity, $L_\text{CR}$, as a function of the emission radius $R$ and power-law index $s_{\rm CR}$ for $\varepsilon_p^{\rm min}=10$~TeV and $\varepsilon_p^{\rm max}=30$~TeV in the minimum hadronic scenario for different values of $\tau_T$. The blue and orange lines have the same meaning as in Figure~\ref{fig:contour-1gev}. The red hatching represents the excluded region for $\xi_B=1$. }
    \label{fig:contour-pp}
\end{figure*}

We consider the standard hadronic scenario, in which neutrinos are produced by $pp$ and $p\gamma$ processes. It has been shown that inelastic $pp$ interactions can give a relevant contribution to the observed neutrinos~\citep[e.g.,][and references therein]{Murase:2022dog}. For hadronuclear interactions, the effective optical depth is given by~\citep{Murase:2019vdl,Murase:2022dog}
\begin{eqnarray}
    f_{pp} &\approx& n_p(\kappa_{pp}\sigma_{pp})R(c/V)
    \approx (\kappa_{pp}\sigma_{pp}/\sigma_T)\zeta_e^{-1}(c/V)\tau_T
    \nonumber \\
    &\simeq& 1.2~\left(\tau_T/0.4\right)\zeta^{-1}_e\left(0.01 c/V\right)
    \label{eq:fpp}
\end{eqnarray}
where $n_p=\tau_T\zeta_e^{-1}\sigma_T^{-1}R^{-1}$ is the target proton density, $\sigma_{pp}\sim4\times10^{-26}~{\rm cm}^2$ is the cross section for $pp$ interactions, $\kappa_{pp}\sim 0.5$ is the proton inelasticity, $\sigma_T\approx6.65\times10^{-25}~{\rm cm}^2$ is the Thomson scattering cross-section, and $\zeta_e$ is the pair loading factor. We explore results for the hadronic scenario at different values of $\tau_T$\footnote{Note that our definition of $\tau_T$ is somewhat different from the corona model that uses $H=R/\sqrt{3}$ for the scale height, in which the Thomson optical depth is given by $\tau_T^{\rm cor}\approx \zeta_e n_p\sigma_TH$.}. This allows us to effectively probe the dominance and influence of this process on the emission spectra and the required cosmic-ray luminosity to explain neutrino observations. Note that $\tau_T=0.1-1$ is expected for the magnetic corona model~\citep{Murase:2019vdl}. We use $\zeta_e=1$ throughout this work. If $\zeta_e$ were larger, our results for $\tau_T=0.1$ and $\tau_T=1$ would look more like those for $\tau_T=10$ and $\tau_T=100$ respectively, since these two parameters are degenerate in the calculation of $n_p$. The effective optical depth for the photomeson production process is estimated to be~\citep{Murase:2019vdl,Murase:2022dog},
\begin{eqnarray}\label{eq:pg-loss}
    f_{p\gamma} &\approx& \eta_{p\gamma} \hat{\sigma}_{p\gamma} R (c /V) \tilde{n}_X \left(\frac{\varepsilon_p}{\varepsilon_{p}^{p\gamma-X}}\right)^{\Gamma_{\rm cor} - 1} \nonumber \\ 
    &\simeq& 1.3~\eta_{p\gamma} \tilde{L}_{\rm cor, 43.3}\mathcal{R}_{1.5}^{-1}\left(0.01 c/V\right)\nonumber\\
    &\times& {\left(\frac{M_{\rm BH}}{10^7M_\odot}\right)}^{-1}{\left(\frac{\varepsilon_p}{\varepsilon_p^{p\gamma-X}}\right)}^{\Gamma_{\rm cor} - 1},
    \label{eq:fpgamma}
\end{eqnarray}
where $\eta_{p\gamma}=2/(1 + \Gamma_{\rm cor})$, $\tilde{n}_X=\tilde{L}_{\rm cor}/(4\pi R^2c\varepsilon_X)$ is the number density of X-ray photons at the reference X-ray energy of $\varepsilon_X=2~{\rm keV}$, $\Gamma_{\rm cor} =2$~\citep{Marinucci:2015fqo}, $\hat{\sigma}_{p\gamma} \sim 0.7\times 10^{-28}\rm~cm^2$, and $\varepsilon_{p}^{p\gamma-X} \simeq 7.9 \times 10^4\rm~GeV~(\varepsilon_{X}/2\rm~keV)^{-1}$. 
In addition, the Bethe-Heitler process is important. The effective optical depth to the Bethe-Heitler pair production process in the disk photon field is estimated to be~\citep{Murase:2019vdl,Murase:2022dog}
\begin{eqnarray}\label{eq:BH-loss}
    f_{\rm BH} &\approx& \tilde{n}_{\rm disk} \hat{\sigma}_{\rm BH} R (c / V) \nonumber \\ 
    &\simeq& 23~\tilde{L}_{\rm disk, 44.7}
    \mathcal{R}_{1.5}^{-1}{\left(\frac{M_{\rm BH}}{10^7M_\odot}\right)}^{-1} \nonumber \\ 
    &\times& {\left(\frac{\varepsilon_{\rm disk}}{31.5\rm~eV}\right)}^{-1} {\left(\frac{V}{0.01 c}\right)}^{-1},
\end{eqnarray}
where $\hat{\sigma}_{\rm BH} \sim 0.8 \times 10^{-30}\rm~cm^2$ and the typical proton energy causing pair production is $\tilde{\varepsilon}_{p}^{\rm BH-disk} \approx 0.5 \bar{\varepsilon}_{\rm BH} m_p c^2 / \varepsilon_{\rm disk} \simeq 1.5 \times 10^5\rm~GeV~(\varepsilon_{\rm disk} / 31.5\rm~eV)^{-1}$ and $\bar{\varepsilon}_{\rm BH} \approx 10\rm~MeV$. The ratio of neutrinos to gamma-ray luminosities is reduced by the Bethe-Heitler process that depletes cosmic-ray protons. Instead, energies of generated lepton pairs are transferred to gamma rays through synchrotron and inverse Compton cascades as well as the two-photon annihilation. Figure~\ref{fig:optical-depth} demonstrates the two-photon annihilation optical depth $\tau_{\gamma\gamma}$, as well as $f_{p\gamma}$, $f_{\rm BH}$, and $f_{pp}$ for a given emission radius.

In the photohadronic scenario, $\sim1-10$~TeV neutrinos are mostly generated due to interactions with X-ray photons from the hot corona. The differential luminosity of neutrinos is given by~\citep{Murase:2019vdl,Murase:2026hrz},
\begin{equation}
\varepsilon_\nu L_{\varepsilon_\nu}\approx \frac{3K}{4(1+K)}{\rm min}[1,f_{\rm mes}]f_{\rm sup}\left[\varepsilon_p L_{\varepsilon_p}\right]_{\varepsilon_p=20\varepsilon_\nu},
    \label{eq:diffLum}
\end{equation}
where $K=1$ for $p\gamma$ and $K=2$ for $pp$ interactions, $\varepsilon_p$ and $\varepsilon_\nu$ are the energies of the proton and the neutrino respectively in the source frame, and $f_{\rm sup}\approx {\rm max}[1,f_{\rm mes}]/f_{\rm BH}$ for $f_{\rm BH}>1$. We focus on cases of $f_{\rm BH}>1$, which is expected in the hadronic scenario for jet-quiet AGNs. In the limit that the Bethe-Heitler process is dominant, the gamma-ray luminosity is written as
\begin{equation}
    \varepsilon_\gamma L_{\varepsilon_\gamma}
    \sim 
    \varepsilon_\gamma\mathcal{G}^s_{\varepsilon_\gamma} 
    \int \frac{d \varepsilon_\gamma}{\varepsilon_\gamma}f_{\rm BH}\left[\varepsilon_p L_{\varepsilon_p}\right]
    \times e^{-\tau_{\gamma\gamma}^{\rm DT}-\tau_{\gamma\gamma}^{\rm EBL}}.
    \label{eq:diffLumBH}
\end{equation}
Here $\varepsilon_\gamma\mathcal{G}^s_{\varepsilon_\gamma}$ represents the source spectrum of cascaded gamma rays~\citep{Murase:2026hrz}, which is normalized by $\int d\varepsilon_\gamma\mathcal{G}^s_{\varepsilon_\gamma}=1$. The two-photon annihilation attenuation factor ${\rm exp}(-\tau_{\gamma\gamma}^{\rm DT}-\tau_{\gamma\gamma}^{\rm EBL})$ may also be incorporated.

\subsection{Multimessenger Spectra}
Figure~\ref{fig:spectrum-plots} shows the cascaded gamma-ray and neutrino spectra for different representations of the standard hadronic scenario, by varying emission radii, values of $\xi_B$, and the minimum energy. The top panels, which correspond to the minimal hadronic scenario, and the narrowband  spectrum can be realized by hard spectra with $s_{\rm CR}<2$ extended from lower-injection energies. These cases mimic magnetically powered corona models such as stochastic acceleration and reconnection acceleration scenarios when the hadronuclear ($pp$) contribution is included~\citep{Murase:2019vdl,Kimura:2019yjo,Eichmann:2022lxh,Murase:2026hrz}. 
The bottom panels, which adopt $s_{\rm CR}=2$ from GeV energies, are representative of models of stochastic~\citep{Lemoine:2023wsw,Lemoine:2024roa} or reconnection~\citep{Kheirandish:2021wkm} acceleration with $s_{\rm CR}\sim2$, and are also broadly relevant to shock acceleration models~\citep{Inoue:2019fil,Inoue:2022yak,Murase:2022dog}. Comparing the bottom two panels, we see that the ratio of neutrino to gamma-ray luminosities is larger when the $pp$ process is included, as expected from Equation~\ref{eq:diffLumBH}. We also find that the GeV gamma-ray luminosity is enhanced when the cascade is inverse-Compton-dominated, e.g., for $\xi_B=0.01$.

\subsection{Multimessenger Constraints}
While~\cite{Das_2024} focused on the minimal hadronic scenario, considering only the purely photohadronic scenario, we extend the analysis by allowing the injected proton spectrum to reach down to 1~GeV, which enables us to assess how low-energy cosmic rays modify the gamma-ray constraints and the required cosmic-ray luminosity. When injected protons are extended down to GeV energies, the gamma-ray constraints become significantly stronger (Figure~\ref{fig:contour-1gev}). For the synchrotron-dominated case ($\xi_B=1$), we obtain $\mathcal{R}\lesssim20$ for the hadronuclear scenario and $\mathcal{R}\lesssim15$ for the photohadronic scenario, while for the inverse-Compton-dominated case ($\xi_B=0.01$), the corresponding limits are $\mathcal{R}\lesssim5$ and $\mathcal{R}\lesssim3$, respectively. At $s_{\rm CR}\sim1$, the cascaded gamma-ray spectra and the resulting constraints are nearly identical to those in the minimal scenario, because the emission is dominated by protons with energies close to $\varepsilon_p^{\rm max}$. However, as $s_{\rm CR}$ increases, the contribution from GeV cosmic rays becomes more important. This enhances the GeV component of the cascaded gamma rays, so that the {\it Fermi}-LAT upper limits are violated for the allowed values of $L_{\rm CR}$ required to match the 1.5 to 15~TeV neutrino flux observed by IceCube. Thus, for softer spectra, the constraints on $R$ for a given value of $\xi_B$ are stronger than in the minimal scenario, where we do not find viable parameter space beyond $s_{\rm CR}\sim2.2$ for $\tau_T=1$ and $s_{\rm CR}\sim2.5$ for the purely photohadronic scenario. Beyond these values, the maximum allowed value of $R$ for $\xi_B\lesssim1$ is smaller than the innermost stable circular orbit (ISCO) of a Schwarzschild black hole which is $\mathcal{R}=3$.

For the minimal scenario, but with hadronuclear contributions, the upper limits from gamma-ray constraints are shown in Figure~\ref{fig:contour-pp} for different values of $\tau_T$ for both the synchrotron-dominated ($\xi_B=1$) and inverse-Compton-dominated ($\xi_B=0.01$) cases. These constraints are calculated for the lower limit of the reported IceCube neutrino luminosity range to be conservative. As $\tau_T$ increases, the $pp$ process becomes more dominant relative to the $p\gamma$ process and the gamma-ray constraints become more relaxed, which is consistent with the findings of~\cite{Murase:2022dog}. 
For $\tau_T=0.1$, where $pp$ interactions are least important, we find $\mathcal{R}\lesssim20$ for $\xi_B=1$ and $\mathcal{R}\lesssim3$ for $\xi_B=0.01$, similar to the photohadronic limits obtained in our previous work. In this regime, inverse-Compton-dominated models are disfavored because the required emission radius would be smaller than the ISCO of a nonrotating black hole. In contrast, for $\tau_T=100$, where the $pp$ process dominates over most of the parameter space, the limits relax to $\mathcal{R}\lesssim70$ for $\xi_B=1$ and $\mathcal{R}\lesssim50$ for $\xi_B=0.01$. For the strongly magnetized case, there is also a second allowed region beyond $\mathcal{R}\sim 9\times10^3$, where the GeV-TeV gamma-ray flux is significantly suppressed through efficient synchrotron cooling of pairs. The intermediate region is excluded due to the violation of the {\it Fermi}-LAT data. This is consistent with the findings of~\cite{Murase:2022dog}. 
The transition radius between $p\gamma$-dominated and $pp$-dominated neutrino production can be estimated by equating the effective optical depths,
\begin{equation}
    \mathcal{R}_{pp=p\gamma}\simeq 33~\eta_{p\gamma}\left(\frac{\tau_T}{0.4}\right)^{-1}\zeta_e
    {\left(\frac{M_{\rm BH}}{10^7M_\odot}\right)}^{-1}\left(\frac{\varepsilon_p}{\varepsilon_p^{p\gamma-X}}\right)^{\Gamma_{\rm cor}-1}.
    \label{eq:Rcrit}
\end{equation}
For $\mathcal{R}\lesssim\mathcal{R}_{pp=p\gamma}$, the $p\gamma$ process due to interactions with coronal X-ray photons is the main source of TeV neutrinos, whereas for $\mathcal{R}\gtrsim\mathcal{R}_{pp=p\gamma}$, the $pp$ process dominates. 

By requiring the satisfaction of gamma-ray constraints for ${\mathcal R}\geq 3$ (i.e., $R\geq R_{\rm ISCO}$) and all possible values of $s_{\rm CR}$, we can also place lower limits on $\xi_B$.~\cite{Das_2024} obtained $\xi_B\gtrsim0.01$ for the minimal photohadronic scenario, which is insensitive in the range of $s_{\rm CR}\lesssim2$. As shown in Figure~\ref{fig:contour-1gev} right, this lower limit is largely unchanged even if the spectrum is extended to GeV energies, which is obvious because low-energy cosmic rays below the threshold do not contribute to gamma rays. However, there should be some contribution from lower-energy cosmic rays, and we have a somewhat stronger limit of $\xi_B\gtrsim0.03$ for $s_{\rm CR}\gtrsim2$. With the inclusion of $pp$ interactions, the limit is alleviated as shown in Figure~\ref{fig:contour-1gev} left. However, for $s_{\rm CR}\gtrsim2$, we obtain $\xi_B\gtrsim0.1$ thanks to enhanced $pp$ contributions. Lower limits on $\xi_B$ can be translated into upper limits on the plasma beta ($\beta$) as~\citep{Murase:2022dog,Murase:2026hrz},
\begin{eqnarray}
    \beta\approx\left(\frac{\tau_T}{6.3\zeta_e\lambda_{\rm Edd}}\right)\xi_B^{-1}\sim6\left(\frac{\tau_T^{\rm cor}}{0.4}\right)\xi_{B,-2}^{-1}\zeta_e^{-1}\lambda_{\rm Edd}^{-1},
    \label{eq:beta}
\end{eqnarray}
where $\tau_T^{\rm cor}=\zeta_e n_p \sigma_T H \approx 0.4$ is motivated by the disk-corona modeling of NGC 1068. Considering effects of $pp$ interactions, our results imply $\beta\lesssim10$, but this limit can be more stringent for $s_{\rm CR}\gtrsim2$.

The minimal hadronic scenario also gives a conservative baseline for the required cosmic-ray power, $L_{\rm CR}$, as found in Figure~\ref{fig:contour-pp}. Introducing hadronuclear interactions can relax the cosmic-ray power requirement relative to the purely photohadronic case, and our results are broadly consistent with the analytical limits on $L_{\rm CR}$ obtained in~\cite{Murase:2022dog}. Together with gamma-ray constraints, allowing for $pp$ interactions can enlarge the viable parameter space, especially at larger values of $\tau_T$.

When GeV protons are included, the required cosmic-ray luminosity is larger, which can be understood through $\mathcal{C}_p$. Because $\mathcal{C}_p$ grows rapidly for softer spectra, the required $L_{\rm CR}$ also increases significantly. For example, if $\varepsilon_p^{\rm min}$ extends to GeV energies, we have $\mathcal{C}_p\simeq 5.3\times10^4$ for $s_{\rm CR}=3.2$, and even for $s_{\rm CR}=2$ we obtain $\mathcal{C}_p\simeq10$, implying $L_{\rm CR}\sim L_{\rm corona}\gtrsim0.1L_{\rm bol}$. Figure~\ref{fig:contour-1gev} shows that the required $L_{\rm CR}$ is nearly identical to that in the minimal scenario at $s_{\rm CR}=1$, but increases rapidly with $s_{\rm CR}$ because GeV cosmic rays contribute to the proton power budget. For $s_{\rm CR}=2$, in the limit $\varepsilon_p^{\rm min}\ll\varepsilon_p^{\rm max}$, Equation~\ref{eq:Cp} gives $\mathcal{C}_p\sim\ln(\varepsilon_p^{\rm max}/\varepsilon_p^{\rm min})$, corresponding to at most a factor of $\sim3$ increase in proton luminosity over the minimal scenario. For softer spectra, however, the required $L_{\rm CR}$ becomes much larger, increasing by many orders of magnitude from $s_{\rm CR}=1$ to $s_{\rm CR}=4$. For $s_{\rm CR}\gtrsim2$, we find $L_{\rm CR}>L_{\rm corona}$ at $\mathcal R\gtrsim3$, which strongly constrains how soft the injected spectrum can be. Note that numerical limits are also somewhat stronger than the simple analytical estimates based on $\mathcal{C}_p$, because the Bethe-Heitler pair production can suppress both the $pp$ and $p\gamma$ neutrinos, thereby increasing the proton power needed to reproduce the neutrino data and making it easier to violate gamma-ray limits.

Overall, our results support the standard hadronic scenario. The purely photohadronic scenario favors a compact, strongly magnetized corona with $\mathcal{R}\sim3-10$. Hadronuclear contributions can either tighten or relax the constraints, but our results favoring strong magnetization for $\tau_T\sim0.1-1$ is consistent with the requirement of magnetically powered corona models. 
Our results may lead to more specific implications for particle acceleration models. For example, accretion shock or failed-wind models proposed in the literature require $s_{\rm CR}\gtrsim2$ and $\xi_B\lesssim0.005$~\citep{Murase:2022dog}, given that the kinetic power is $L_{\rm kin}\approx (1/2)\eta_{\rm kin}\dot{M}V^2$, where $\eta_{\rm kin}$ is the energy fraction carried by inflows or outflows, and the energy fraction carried by magnetic fields is typically $\epsilon_B\sim{10}^{-3}-{10}^{-2}$. Our constraints are in tension with the fiducial values of $s_{\rm CR}$ and $\xi_B$. In particular, for $\xi_B=0.01$, there is essentially no viable parameter space for $\mathcal{R}\gtrsim3$ for $\tau_T=0.1-1$ and in the purely photohadronic case. In addition, if X rays are supplied by the shock dissipation, $L_{\rm CR}< L_{\rm kin}\sim L_{\rm corona}$ is expected, which also leads to a tension with our constraint, $L_{\rm CR}\gtrsim L_{\rm corona}$.  
However, shock models may still be viable if the emission region is strongly magnetized and/or extremely dense (e.g., $\tau_T\sim100$). 
On the other hand, magnetically powered corona models have larger viable parameter space, being consistent with our multimessenger constraints. For example, the turbulent corona model with stochastic acceleration is expected to have $\beta\lesssim1-3$, corresponding to $\xi_B\gtrsim0.03$, and reconnection-based models can imply even larger values of $\xi_B$. 
These values are consistent with our lower limits on $\xi_B$. They may have hard spectra of $s_{\rm CR}\lesssim2$, in which $L_{\rm CR}\lesssim L_{\rm corona}$ can be satisfied.

\section{Beta Decay scenario}
\begin{figure*}
    \centering    
    \includegraphics[width=\linewidth]{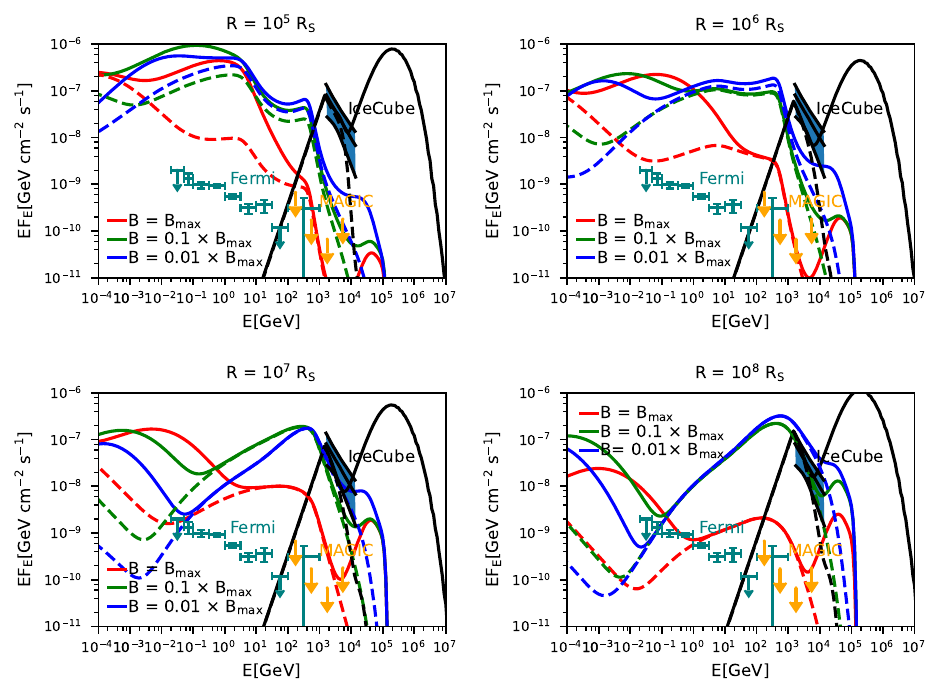}
    \caption{Cascaded gamma-ray spectra (colored lines) and all-flavor neutrino spectra (black lines) for different values of $R$ and magnetic field strength in the decay scenario. The magnetic field strengths used are computed in terms of $B_{\rm max}$ as defined in Equation~\ref{eq:Bmax}. The solid lines correspond to cascades computed with the photomeson contribution and the dashed lines correspond to cascades neglecting the photomeson contribution. Attenuation due to CMB and EBL~\citep{Gilmore:2011ks} is considered. All panels are made using $s_{\rm CR}$ = 3.2, $\varepsilon_{A}^{\rm min}$ =5~PeV and $\varepsilon_{A}^{\rm max}$=15~PeV for the injected helium nuclei spectrum.}
    \label{fig:result-decay}
\end{figure*}
\begin{figure}
    \centering    
    \includegraphics[width=0.5\textwidth]{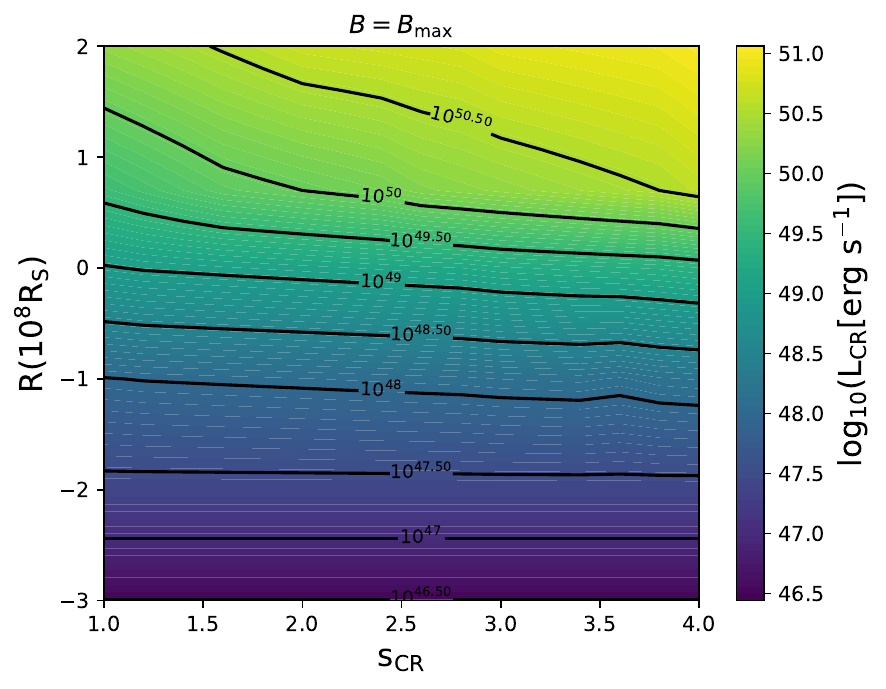}
    \caption{Required minimum cosmic-ray helium luminosity, $L_{\rm CR}$, as a function of $R$ and $s_{\rm CR}$ in the beta decay scenario for $B=B_{\rm max}$ as defined in Equation~\ref{eq:Bmax}. We find that $L_{\rm CR}$ exceeds $L_{\rm bol}\sim L_{\rm Edd}$ in the entire parameter space. Gamma-ray constraints from the {\it Fermi}-LAT data cannot be shown because the entire parameter space is excluded by the {\it Fermi}-LAT data (see Figure~\ref{fig:result-decay}).}
    % \vspace{-5mm}
    \label{fig:contour-decay}
\end{figure}

In the beta decay scenario, nuclei accelerated in the source undergo photodisintegration interactions with ambient photons, producing neutrons. These neutrons eventually decay via beta decay, yielding electron antineutrinos as well as secondary electrons. This has been studied especially in the context of extragalactic ultrahigh-energy cosmic-ray accelerators~\citep{Murase:2010va,Aharonian:2010te,Zhang:2023ewt}, and it has recently been applied to NGC 1068~\citep{Yasuda_betaDecay}. 

\cite{Das_2024} explored this beta-decay scenario, considering the infrared field from the dust torus, and showed that it is excluded as the dominant mechanism for neutrinos from NGC 1068. Although $B=1~\mu$G is adopted as a default value,~\cite{Das_2024} also considered an equipartition magnetic field introduced as $B_{\rm eq}=\sqrt{(8\pi L_{\rm DT})/(4\pi R^2 c)}\simeq0.24~{\rm G}~{(R/R_{\rm DT})}^{-1}{(0.1~{\rm pc}/R_{\rm DT})}L_{\rm DT,43.9}^{1/2}$. They further examined a stronger magnetic field, $B=300~\mu{\rm G}$, in which the pairs may predominantly cool via synchrotron losses at large radii, and concluded that their main conclusions remain unchanged.~\cite{Yasuda_betaDecay} considered cases in which the pairs mainly lose energy through inverse-Compton emission in the first version of their preprint. Later, they considered secondary synchrotron emission as well as the infrared radiation field, and argued that the beta decay scenario can satisfy the {\it Fermi}-LAT constraints if sufficiently strong magnetic fields are adopted. Although~\cite{Das_2024} had actually explored such a strong field regime, including cases where synchrotron cooling dominates based on the equipartition magnetic field, we here present the results more explicitly. 

The jet power ($L_{\rm jet}$) can be expressed in terms of the mass accretion rate ($\dot{M}$) and the Eddington luminosity as follows:
\begin{eqnarray}
P_{\rm jet}=\eta_{\rm jet}\dot{M}c^2=\lambda_{\rm Edd}\left(\frac{\eta_{\rm jet}}{\eta_{\rm rad}}\right)L_{\rm Edd},
\end{eqnarray}
where $\eta_{\rm jet}$ is the jet efficiency and $\eta_{\rm rad}=L_{\rm bol}/(\dot{M}c^2)$ is the radiative efficiency. Observations of jet-loud AGNs suggest $P_{\rm jet}/L_{\rm Edd}\sim0.01-1$ at $\lambda_{\rm Edd}\sim1$~\citep{Fan:2019ojg}, and it is more likely that jet-quiet AGNs such as NGC 1068 have smaller values~\citep{Salvatore:2023zmf}. Thus, it is conservative to require $P_{\rm jet}\leq L_{\rm Edd}$, and we parameterize the magnetic field strength in terms of the maximum allowed magnetic field, $B_\text{max}$, which can be calculated as follows,
\begin{eqnarray}
    B_\text{max}&=&\sqrt{2L_\text{Edd}/(R^2c)} \nonumber \\ 
    &\simeq&1.0~\left(\frac{M_{\rm BH}}{10^7M_{\odot}}\right)^{1/2} \mathcal{R}_5^{-1}~{\rm G}.
    \label{eq:Bmax}
\end{eqnarray}
We explore the results for $B = b B_\text{max}$, where $b\in [0.01, 0.1, 1]$ in this work. 
%For example, the maximum allowed magnetic field strength at $R=0.8$~pc is $\sim0.12~\rm~G$ as calculated using Equation.~\ref{eq:Bmax}. 
%However,~\cite{Yasuda_betaDecay} calculate estimates using a field strength of 3~G at 0.8~pc which would require the source to be super-Eddington. We know that this is not the case for NGC 1068.

We assume an injected spectrum of cosmic-ray helium nuclei with the same functional form described in Equation~\ref{eq:injCR}. The minimum energy of injected helium ions, $\varepsilon^{\rm min}_A$, is set to 5~PeV and the cutoff energy $\varepsilon^{\rm max}_A$ is set to 15~PeV. This choice is conservative, and our results are not affected if $\varepsilon^{\rm max}_A$ is set to 100~PeV. 
We vary the injected cosmic-ray power-law index $s_{\rm CR}$ from 1 to 4 just as we have done in the hadronic scenario. 
Radio observations indicate a sub-relativistic outflow with $V\lesssim0.1c$~\citep{2023ApJ...953...87F}. We set the dynamical velocity to $V=0.1c$ which is the same as the value used in \cite{Das_2024}. We also checked the results for $V=0.01c$ and our conclusions remain unchanged. 

We consider the photodisintegration, photomeson production, Bethe-Heitler pair production, and neutron decay processes. The effective photodisintegration cross section for $A\leq4$ (including inelasticity with $1/A$) is estimated by $\hat{\sigma}_{\rm dis}\approx 4.3\times{10}^{-28}~{\rm cm}^2~{(A/4)}^{-2.433}$~\citep{Karakula:1994nv,Murase:2010va}. 
For energies of interest, optical photons are relevant target for nuclei, and the effective optical depth to photodisintegration for helium nuclei is~\citep{Das_2024}
\begin{eqnarray}
f_{\rm dis}&\approx&\tilde{n}_{\rm disk}\hat{\sigma}_{\rm dis}R (c/V)
\nonumber \\
&\simeq& 0.38~\tilde{L}_{\rm disk, 44.7} {\mathcal R}_5^{-1}{\left(\frac{M_{\rm BH}}{10^7M_\odot}\right)}^{-1}\nonumber\\
&\times& \left(\frac{\varepsilon_{\rm disk}}{31.5\rm~eV}\right)^{-1}  \left(\frac{V}{0.1 c}\right)^{-1},
\end{eqnarray}
which can be $\sim 1$ at sufficiently small radii. Note that the fraction of energy carried by nucleons is only $1/A$ per collision, but all the details are fully taken into account numerically thanks to {\sc AMES}~\citep[e.g.,][]{Zhang:2023ewt}. The average energy of neutrinos from neutron decay is $\sim0.48$~MeV in the neutron rest frame, so the neutrino energy is $\varepsilon_\nu \approx 0.48~{\rm MeV}~\gamma_{n}\simeq 0.64~{\rm TeV}~(\varepsilon_{\rm A}/5~{\rm PeV})$, where $\gamma_n\approx\gamma_{A}\approx\varepsilon_{\rm A}/(Am_pc^2)\simeq1.3\times10^6~(\varepsilon_{\rm A}/5~{\rm PeV})$~\citep[e.g.,][]{Murase:2010va}. The differential all-flavor neutrino luminosity can be analytically expressed in terms of the differential neutron and helium luminosities as
\begin{eqnarray}
\varepsilon_\nu L_{\varepsilon_{\nu}}
&\approx& \frac{1}{2000} \varepsilon_n L_{\varepsilon_n}\nonumber\\
&\approx & \frac{1}{2000}\frac{1}{2}{\rm min}[1,f_{\rm dis}]\varepsilon_A L_{\varepsilon_A},
\label{eq:nuLumBeta}
\end{eqnarray}
where $\varepsilon_n (dL_{\varepsilon_n}/d\varepsilon_n)$ is the differential neutron luminosity and the energy fraction carried by neutrinos is $0.48~{\rm MeV}/(m_nc^2)\sim1/(2000)$. The Bethe-Heitler pair production process does not deplete cosmic-ray nuclei, but its electromagnetic 
contribution cannot be ignored~\citep{Das_2024}. The effective optical depth is estimated to be
\begin{eqnarray}\label{eq:BH-loss-A}
f^{\rm A}_{\rm BH} &\approx& \tilde{n}_{\rm disk}(Z^2/A)\hat{\sigma}_{\rm BH} R (c / V) \nonumber \\ 
&\simeq& 7.0\times10^{-4}~\tilde{L}_{\rm disk, 44.7}
    \mathcal{R}_5^{-1}{\left(\frac{M_{\rm BH}}{10^7M_\odot}\right)}^{-1} \nonumber \\ 
&\times& {\left(\frac{\varepsilon_{\rm disk}}{31.5\rm~eV}\right)}^{-1} {\left(\frac{V}{0.1 c}\right)}^{-1},
\end{eqnarray}
where $(Z^2/A)\hat{\sigma}_{\rm BH} \sim 0.8 \times 10^{-30}\rm~cm^2$ is the attenuation cross section for helium, and the typical energy of a helium ion causing pair production is $\tilde{\varepsilon}_{\rm He}^{\rm BH-disk} \approx 0.5 \bar{\varepsilon}_{\rm BH} m_{\rm He} c^2 / \varepsilon_{\rm disk}\simeq 6.0 \times 10^5\rm~GeV~(\varepsilon_{\rm disk} / 31.5\rm~eV)^{-1}$ and $\bar{\varepsilon}_{\rm BH} \approx 10\rm~MeV$, where $\gamma_{\rm BH}\approx\gamma_A$. The synchrotron luminosity is $\varepsilon_\gamma L_{\varepsilon_\gamma}^{\rm syn}\sim [1/2/(1+Y)]f_{\rm BH}\varepsilon_AL_{\varepsilon_A}$, which has a peak at $\varepsilon_{\gamma,\rm syn}^{\rm pk}\approx 1.5\hbar(eB/m_e c)\gamma_{\rm BH}^2 \simeq 3.0~{\rm GeV}~b (M/10^7M_{\odot})^{-1/2}\left(\varepsilon_{\rm A}/5~{\rm PeV}\right)^2\mathcal{R}_5^{-1}$.
The inverse-Compton luminosity is $\varepsilon_\gamma L_{\varepsilon_\gamma}^{\rm IC}\sim [Y/2/(1+Y)]f_{\rm BH}\varepsilon_AL_{\varepsilon_A}$, which would peak around $\varepsilon^{\rm pk}_{\gamma \rm, IC}\approx(4/3)\gamma_{\rm BH}^2\varepsilon_{\rm DT}\simeq540~{\rm GeV}~(\varepsilon_{\rm DT}/0.24~{\rm eV})\left(\varepsilon_{\rm A}/5~{\rm PeV}\right)^2$. Here, $\varepsilon_{\rm DT}\approx2.82k_BT_{\rm DT}$ and $Y$ is the Compton Y-parameter determined by the magnetic field strength. 
Then, in the limit that the contribution from the photomeson production is negligible, the differential gamma-ray luminosity is estimated by
\begin{eqnarray}
\varepsilon_\gamma L_{\varepsilon_\gamma}\sim 
    \varepsilon_\gamma\mathcal{G}^s_{\varepsilon_\gamma}
    \int \frac{d\varepsilon_\gamma}{\varepsilon_\gamma} f^A_{\rm BH}[\varepsilon_A  L_{\varepsilon_A}]
    \times e^{-\tau_{\gamma\gamma}^{\rm EBL}},
    \label{eq:diffLumBeta}
\end{eqnarray}
where $\mathcal{G}^s_{\varepsilon_\gamma}$ is introduced as in the hadronic scenario, but details are different due to different target radiation fields and magnetic field strengths. The extra two-photon annihilation attenuation factor, ${\rm exp}(-\tau_{\gamma\gamma}^{\rm EBL})$, is also implemented. 

\subsection{Multimessenger Spectra}
Figure~\ref{fig:result-decay} shows neutrino and cascaded gamma-ray spectra for different values of the emission radii and magnetic field strengths. While the electromagnetic and neutrino luminosities are comparable at the injection, the fate of the electromagnetic energy depends on the ratio of the magnetic energy density to the radiation energy density. This is especially apparent in the gamma-ray spectrum for $B=B_{\rm max}$ in the bottom right panel of Figure~\ref{fig:result-decay}, since the synchrotron cooling significantly diminishes the GeV gamma-ray flux compared to those for weaker magnetic field strengths while leaving an excess at MeV energies. The inverse-Compton peak at sub-TeV energies is prominent in the spectra for $B=0.01B_{\rm max}$ and $B=0.1B_{\rm max}$ in all panels of Figure~\ref{fig:result-decay}. For $\mathcal{R}\sim10^5$, the attenuation due to photons from the dust torus is significant at sub-TeV energies, which can readily be seen in the top left panel of Figure~\ref{fig:result-decay}. 

The Bethe-Heitler pair production process is unavoidable because the threshold for pair production, $2m_ec^2$, is lower than that for nuclear disintegration, and its contribution to the electromagnetic energy is significant as seen from Equation~\ref{eq:diffLumBeta}. We should note that a similar efficiency issue also occurs for deexcitation gamma rays even though their energy range is different from that of cascaded gamma rays from the Bethe-Heitler pairs and beta decay electrons~\citep{Murase:2010va,Aharonian:2010te,Zhang:2023ewt}.

\cite{Yasuda_betaDecay} assumes strong magnetic fields corresponding to $P_{\rm jet}>L_{\rm Edd}$. They also adopt a box ($\delta$-function-like) approximation to evaluate the spectrum of pairs from the Bethe-Heitler pair production process, i.e., a simplified treatment in which the electron-positron spectrum is distributed in a very narrow energy interval. As a result, the gamma-ray spectrum has a narrower peak at very high energies, in which inverse Compton photons are significantly attenuated by the two-photon annihilation with infrared photons from the dust torus as long as $\tau_{\gamma\gamma}>1$. However, such an approximation may not appropriately describe GeV gamma-ray spectra in our setup~\cite[see, e.g.,][as a demonstration of the importance of detailed spectra]{Zhang:2023ewt,Petropoulou:2026xrr}. Details of the spectrum of Bethe-Heitler pairs broaden resulting gamma-ray spectra, which gives a reason why our detailed cascaded gamma-ray spectra always overshoot the {\it Fermi}-LAT data.  

Last but not least, we must account for the primary gamma-ray component from $\pi^0$ decay produced by the photomeson production. The threshold energy for the pion production is higher, but due to the higher efficiency~\citep[see also][]{Murase:2010va}, the contribution from the photomeson production is not negligible in general, as can be seen from Figure~\ref{fig:result-decay}.

\subsection{Multimessenger Constraints}
The gamma-ray spectrum in Figure~\ref{fig:result-decay} largely exceeds the upper limits from {\it Fermi}-LAT and MAGIC data, even for the maximum permissible magnetic field strength, $B_{\rm max}$. As in~\cite{Das_2024}, the beta decay scenario for neutrinos from NGC 1068 is excluded, at least for the parameter space explored in this work.    

The required values of $L_{\rm CR}$ are shown in Figure~\ref{fig:contour-decay} for the entire parameter space explored. The contours are nearly identical to the results presented in~\cite{Das_2024}. The required $L_{\rm CR}$ is greater than the Eddington luminosity in the entire parameter space, $\gtrsim3\times 10^{46} {\rm~erg~s^{-1}}$. It increases almost linearly with the emission radius while being very weakly dependent on $s_{\rm CR}$, except at very large radii ($R\gtrsim 10^9 R_{\rm S}$). The required $L_{\rm CR}$ is also independent of our choice of magnetic field strengths just as found in~\cite{Das_2024}.

The neutrino spectrum has two peaks, with the lower-energy peak coming from beta decay of neutrons and the higher-energy peak coming from photomeson production. However, the latter peak is beyond the range of neutrinos detected by IceCube. This may leave room for further constraints from future detections by high-energy neutrino detectors. 

Equation~\ref{eq:nuLumBeta} gives the differential luminosity of the neutrino spectrum in terms of that of the injected helium spectrum. From the ratios, we can see that regardless of the effective optical depth, the cosmic-ray power exceeds the neutrino luminosity by a factor of $\sim{10}^4$. Thus, we have
\begin{equation}
L_{\rm CR} \gtrsim {10}^4 L_\nu \sim {10}^{46}~{\rm erg}~{\rm s}^{-1} \gg L_{\rm Edd},  
\end{equation}
which agrees with our numerical simulations shown in Figure~\ref{fig:contour-decay}. The situation is worse if we consider more realistic spectra with the cosmic-ray spectrum extended to $\varepsilon^{\rm min}_A\sim2-10$~GeV rather than $\varepsilon^{\rm min}_A=5$~PeV. Thus, from the energetics point of view, we also conclude that the beta decay scenario is challenging as the dominant origin of NGC 1068 neutrinos.

\section{Summary and Discussion}\label{sec:summary}

In this study, based on the latest multimessenger observations including IceCube and {\it Fermi}-LAT data, we provided constraints on the cosmic-ray luminosity and emission radius that are compatible with the Seyfert II galaxy NGC 1068 by extending our analysis presented in~\cite{Das_2024}.
Our results are summarized as follows.

\begin{enumerate}
    \renewcommand{\labelenumi}{(\roman{enumi})}
    \item The hadronuclear scenario allows for somewhat larger emission radii than the photohadronic scenario. The higher the coronal density and efficiency of the $pp$ process, the larger the allowed emission radius. Just as we see for the photohadronic scenario, the constraints for low-$\beta$ plasma are weaker than those for high-$\beta$ plasma. Indeed, for $\tau_T\sim0.1-1$, which is expected in X-ray coronae, we found ${\mathcal R}\lesssim20-30$ for $\xi_B=1$ and ${\mathcal R}\lesssim3-5$ for $\xi_B=0.01$, respectively. The corresponding limits in $p\gamma$ scenarios are ${\mathcal R}\lesssim15$ and ${\mathcal R}\lesssim3$, respectively~\citep{Das_2024}, supporting a compact, strongly magnetized corona with ${\mathcal R}\sim3-10$ that is expected in high Eddington ratio systems~\citep{Murase:2026hrz,Carpio:2026xkf}. 
    In more extreme cases with $\tau_T\sim1-100$ (in which the neutrino emission region should not be the X-ray corona), we found that the constraints are relaxed as ${\mathcal R}\lesssim30-70$ for $\xi_B=1$ and ${\mathcal R}\lesssim5-50$ for $\xi_B=0.01$, respectively, being consistent with findings of~\citet{Murase:2022dog}. Though challenging in terms of required physical conditions, we also confirmed possible parameter space with ${\mathcal R}\gtrsim9000$, as found in~\citet{Murase:2022dog}. Our results encourage further investigations into $pp$ neutrino contributions. 
    \item While the assumption of $\varepsilon_p^{\rm min}=10$~TeV is useful for the purpose of obtaining the minimum cosmic-ray power, it is more reasonable to expect that the cosmic-ray spectrum is extended down to GeV energies. For $\varepsilon_p^{\rm min}=1$~GeV, we found that the required values of $L_{\rm CR}$ exceed the total X-ray luminosity, $L_{\rm corona}$, for $s_{\rm CR}\gtrsim2$, and no parameter space exists for $\xi_B\lesssim0.03$. This may strongly disfavor the specific shock model attributing the coronal X-ray emission to shock dissipation, which typically predicts $L_{\rm CR}< L_{\rm corona}$ and $\xi_B\lesssim0.005$. 
    However, shock acceleration itself remains viable if $L_{\rm CR}\sim L_{\rm corona}$ and $\xi_B\gtrsim0.03-1$, which could be realized when shocks occur through magnetic dissipation in magnetically powered coronae~\citep{Murase:2022dog}. 
    For $\tau_T\sim0.1-1$, which is the case if neutrinos come from the X-ray corona, we found that $L_{\rm CR}< L_{\rm corona}$ requires hard cosmic-ray spectra with $s_{\rm CR}\lesssim2$, supporting magnetically powered corona models involving strong turbulence and magnetic reconnections~\citep{Murase:2019vdl,Kheirandish:2021wkm,Fiorillo:2023dts,Lemoine:2023wsw,Lemoine:2024roa}. 
    While this work focuses on NGC 1068, applications to other AGNs such as NGC 4151 and NGC 7469~\citep{Murase:2023ccp,Saurenhaus:2025ysu,Yang:2025lmb,Carpio:2026xkf,Testagrossa:2026jcs,Eichmann:2026kvj} would also be of interest. 
    \item The beta decay scenario violates both energetics and gamma-ray constraints even if the magnetic field is as strong as the maximum values allowed by the Eddington luminosity. The suppression of the inverse-Compton cascade by strong synchrotron cooling of pairs is not significant enough to make the observed gamma-ray flux below the upper limits set by {\it Fermi}-LAT and MAGIC data. We explicitly showed that the conclusions from~\cite{Das_2024} remain unchanged, and the beta decay scenario is unlikely as the dominant origin of NGC 1068 neutrinos. However, we note that such a setup can be relevant in other sources~\citep[e.g.,][]{Pe'er:2009rc,Zhang:2023ewt}, and applications to AGN jets could remain interesting.  
\end{enumerate}

In reality, there are multiple regions in which gamma-ray emission occurs. While we work with a single-zone model in this work, our constraints are conservative in the sense that we study the implications of the dominant neutrino emission region. The presence of additional gamma-ray emitting regions would further reduce the gamma-ray budget available for neutrino production and make the constraints on the neutrino emitting region tighter. Note that we do not attempt to fit the gamma-ray data. Rather, we require that the gamma-ray flux associated with the neutrino flux not violate the {\it Fermi}-LAT data and MAGIC upper limits. We should also note that the outer emission regions will be less efficient in neutrino production than the innermost region since the target photon density decreases as the distance from the SMBH increases. Our results would be useful for future investigations that aim to bridge the gap between phenomenological acceleration models~\citep[e.g.,][]{Murase:2019vdl,Kheirandish:2021wkm,Murase:2026hrz} and first-principles numerical simulations. Several numerical studies of dynamic astrophysical environments have been conducted at multiscales, including particle-in-cell (PIC) or equivalent kinetic simulations~\citep[e.g.,][]{Comisso:2018kuh,Zhdankin:2018lhq, Mbarek:2022nat,Sridhar:2022ojr,PhysRevD.109.L101306, Groselj:2026nix}, magnetohydrodynamic (MHD) with test-particle injection~\citep[e.g.,][]{Kimura:2016fjx, Kimura:2018clk, Sun:2021ods}, and hybrid PIC-MHD~\citep[e.g.,][]{Liu:2025gxf} simulations. Using kinetic (PIC or hybrid PIC-MHD) simulations, particle injection spectra, acceleration efficiencies, and diffusion coefficients may be inferred, which can be confronted with our general constraints derived from multimessenger observables. 

AGNs have been discussed as laboratories for testing physics beyond the Standard Model, and our constraints will be useful for probing neutrino-dark matter interactions and nucleon-dark matter interactions through cosmic-ray cooling and boosted dark matter \cite{Herrera:2023nww, Gustafson:2024aom, Gustafson:2025dff}.

%%%%%%%%%%%%%%%%%%%%%%%%%%%%%%%%%%%%%%%%%%%%%%%%%%
%%%%%%%%%%%%%%%%%%%%%%%%%%%%%%%%%%%%%%%%%%%%%%%%%%

\begin{acknowledgements}
We thank Eduardo Guti\'{e}rrez, Eduardo Munguia Gonzalez, and Yujia Wei for the useful discussions. 
K.M. deeply thanks Yoshiyuki Inoue and Nobuyuki Sakai for discussions that clarifies the treatment of the Bethe-Heitler pair production used in~\citet{Yasuda_betaDecay}. 
The research by A.D. and K.M. is supported by the NSF Grant No.~AST-2308021. K.M. was previously supported by the NSF Grants No.~AST-2108466 and No.~AST-2108467. 
%The research by B.T.Z. and K.M. is supported by KAKENHI No.~20H05852.
K.M. also acknowledges Mamoru Yanagisawa for his generous donation and continuous support. 

We recently became aware of~\cite{LeBihan:2026rqq}. While their methodology is different from our model-independent phenomenological framework for particle injection and acceleration, the results are consistent with each other.
\end{acknowledgements}
\clearpage
\appendix
\section{Impact of Bethe--Heitler Pair Production}
\label{app:bh_importance}

In this appendix, we illustrate the impact of Bethe-Heitler pair production on the neutrino and electromagnetic cascade spectra. Bethe-Heitler pair production is an unavoidable energy-loss channel for cosmic rays propagating in the intense radiation fields expected in the compact core of NGC~1068. Although this process does not directly produce neutrinos, it transfers part of the cosmic-ray energy into electron-positron pairs, which subsequently contribute to the electromagnetic cascade. It can therefore modify the partition of the injected cosmic-ray power between neutrinos and photons.

\begin{figure*}[htbp]
    \centering

    \begin{minipage}[t]{0.485\textwidth}
        \centering
        \includegraphics[width=\linewidth]{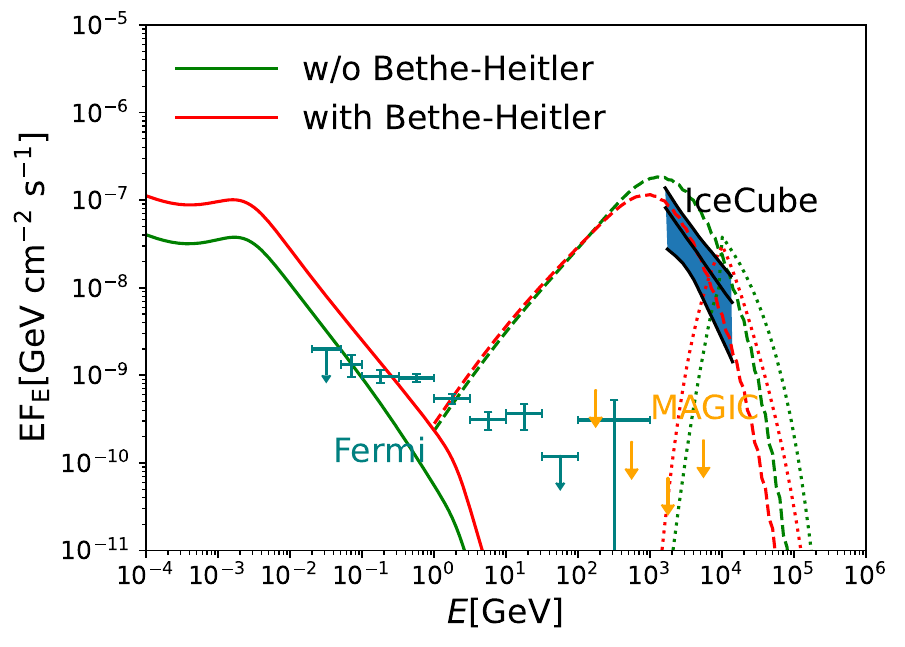}
        \vspace{0.5ex}

        \textbf{(a)} $pp+p\gamma$
    \end{minipage}
    \hfill
    \begin{minipage}[t]{0.485\textwidth}
        \centering
        \includegraphics[width=\linewidth]{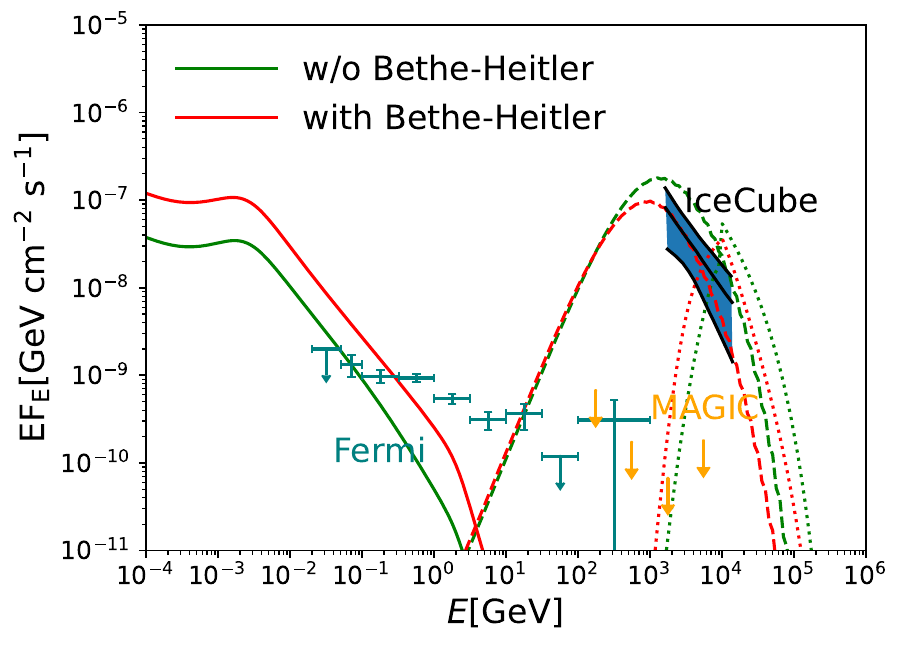}
        \vspace{0.5ex}

        \textbf{(b)} $p\gamma$ only
    \end{minipage}

    \caption{
   Comparison of the neutrino and cascaded electromagnetic spectra obtained with and without Bethe-Heitler pair production. Solid, dashed and dotted lines correspond to the electromagnetic, neutrino, and proton spectra respectively. The results shown are for $R=3R_S$, $s_{\rm CR}=2$, and $\xi_B=0.01$ in the \emph{minimal} scenario ($\varepsilon_p^{\rm min}=10~$TeV).
    Left: hadronuclear and photohadronic ($pp+p\gamma$) interactions with $\tau_T=0.4$.
    Right: photohadronic ($p\gamma$) interactions only.}
    \label{fig:app_bh_onoff}
\end{figure*}

Figure~\ref{fig:app_bh_onoff} compares the spectra obtained with and without Bethe-Heitler pair production for a representative compact, radiation-dense case with
$R=3R_S$, $s_{\rm CR}=2$, and $\xi_B=0.01$.
We show the results for when both hadronuclear ($pp$) and photohadronic ($p\gamma$) processes are considered, as well as when only the photohadronic process is considered.
In the $p\gamma$-dominated case, including Bethe-Heitler pair production visibly increases the electromagnetic cascade component while reducing the neutrino output, since part of the cosmic-ray energy is diverted into electron-positron pairs rather than photomeson production.
When both $pp$ and $p\gamma$ processes are considered, the resulting electromagnetic spectrum escaping the source is relatively less strongly modified because of the increased electromagnetic contribution from $pp$ interactions, especially in the sub-TeV range where it is the dominant process. Nevertheless, the neutrino spectrum and the internal energy partition are still affected.

This comparison demonstrates that Bethe-Heitler pair production is not merely a subdominant radiative correction in compact coronal models of NGC~1068. 
Rather, it is part of the self-consistent cosmic-ray energy budget. The final qualitative constraints can remain similar to previous studies that did not explicitly emphasize this channel \citep[e.g., ][]{Blanco:2023dfp}, but the inclusion of Bethe-Heitler pair production can alter the detailed neutrino-to-gamma-ray energy partition and shift the corresponding parameter-space constraints.

\bibliography{kmurase}
\bibliographystyle{aasjournalv7}

\end{document}